\documentclass[aps,pre,twocolumn,groupedaddress]{revtex4}

\begin{document}

\title{Scaling limit of vicious walks
and two-matrix model}

\author{Makoto Katori}
\email[]{katori@phys.chuo-u.ac.jp}
\altaffiliation{On leave from, Department of Physics,
Faculty of Science and Engineering,
Chuo University, Kasuga, Bunkyo-ku, Tokyo 112-8551, Japan}
\affiliation{
University of Oxford, 
Department of Physics--Theoretical Physics, 
1 Keble Road, Oxford OX1 3NP, United Kingdom}
\author{Hideki Tanemura}
\email[]{tanemura@math.s.chiba-u.ac.jp }
\affiliation{
Department of Mathematics and Informatics,
Faculty of Science, Chiba University, 1-33 Yayoi-cho, Inage-ku,
Chiba 263-8522, Japan}

\date{\today}

\begin{abstract}
We consider the diffusion scaling limit of the one-dimensional
vicious walker model of Fisher 
and derive a system of nonintersecting Brownian motions. 
The spatial distribution of $N$ particles is studied and it
is described by use of the probability density
function of eigenvalues of $N \times N$ Gaussian random matrices.
The particle distribution depends on the ratio of the
observation time $t$ and the time interval $T$ in which the
nonintersecting condition is imposed.
As $t/T$ is going on from 0 to 1, there occurs a transition 
of distribution, which is identified with the transition 
observed in the two-matrix model of Pandey and Mehta.
Despite of the absence of matrix structure in the original
vicious walker model, in the diffusion scaling limit, 
accumulation of contact repulsive interactions realizes 
the correlated distribution of eigenvalues in 
the multimatrix model as the particle distribution.\\
\nonumber PACS numbers: 05.40.-a, 05.50.+q, 02.50.Ey
\end{abstract}

\pacs{05.40.-a, 05.50.+q, 02.50.Ey}

\maketitle

\section{INTRODUCTION \label{sec:Intro}}

The vicious walker models, in which random walkers walk
without intersecting with any others in a given time interval,
were introduced by Michael Fisher and applications of the models
to various wetting and melting phenomena were described in 
his Boltzmann medal lecture \cite{Fis84}.
Recently, using the standard one-to-one correspondence between
walks and Young tableaux, Guttmann {\it et al.}\cite{GOV98}
and Krattenthaler {\it et al.}\cite{KGV00} showed that exact
formulae for total numbers of one-dimensional vicious walks,
some of which were conjectured in previous papers 
\cite{Fis84,HF84,AME91,EG95}, are derived following the theory of
symmetric functions associated with Young diagrams 
\cite{Mac95,Ful97,Sta99} or the representation theory
of classical groups \cite{FH91}. Important analogies between
the ensembles of Young tableaux and those of Gaussian random
matrices were reported by Johansson \cite{Joh00}, and then
Baik \cite{Bai00} and Nagao and Forrester \cite{NF02,Fold} studied
the vicious walker models using the 
random matrix theory \cite{Meh91,Dei00}.

The purpose of the present paper is to demonstrate more
explicit relations among the vicious walker model, 
the symmetric function called the Schur function and the
Gaussian ensembles of random matrices by considering
the diffusion scaling limit of the one-dimensional vicious
walks. Since each random walk converges to a Brownian 
motion in the scaling limit, the limit process of $N$
vicious walkers will be a system of $N$ {\it nonintersecting
Brownian motions} \cite{Spohn}. In order to enumerate
all possible nonintersecting paths of
walkers realized on a spatio-temporal plane,
we use the so-called Lindstr\"om-Gessel-Viennot formula
\cite{Lin73,GV85,Stem90}, which leads us to a useful
determinantal expression for the transition probability
density of nonintersecting Brownian motions.
We found that its initial-configuration dependence 
can be generally described by using the Schur 
function and the well-known properties of this function 
enable us to define the nonintersecting Brownian
motions in which {\it all particles start from a single
position}. Because the nonintersecting condition will be
imposed for a given time interval, say $T$, all the particles
are immediately disunited from the initial point,
and then they walk randomly 
keeping the nonintersecting condition. We have studied
the time dependence of the spatial distribution
of particle positions. We report in this paper that
the position distribution of $N$ nonintersecting
Brownian motions can be identified with the distribution
of eigenvalues of $N \times N$ complex hermitian matrix $H$
coupled to a real symmetric matrix $A$, in which
$H$ and $A$ are randomly chosen from the Gaussian ensembles.
Such a {\it two-matrix model} was studied by Pandey and Mehta
\cite{PM83,MP83}, in which one parameter was introduced to
control the coupling strength between two matrices.
We will show that the time dependence of our process
can be expressed by the parameter dependence of
Pandey-Mehta's two-matrix model.

Here we consider the probability density function
of $N$ real variables $\{x_{1}, \cdots, x_{N}\}$
with a real parameter $\beta \geq 0$,
\begin{eqnarray}
 P_{\beta}(x_{1}, \cdots, x_{N}) 
&=& C e^{-\beta \sum x_{j}^2/2} 
\prod_{1 \leq j < k \leq N} 
|x_{j}-x_{k}|^{\beta} \nonumber\\
&=& C \exp (-\beta W(\{x_{j}\}) )
\label{eqn:Pbeta}
\end{eqnarray}
with
\begin{equation}
W(\{x_{j}\})=\frac{1}{2} \sum_{j=1}^{N} x_{j}^2
-\sum_{1 \leq j < k \leq N} \ln 
|x_{j}-x_{k}|,
\label{eqn:W}
\end{equation}
where $C$ is a normalization constant.
It is known that (\ref{eqn:Pbeta}) with $\beta=1,2$ and 4
describe the distributions of eigenvalues of random
matrices in the Gaussian orthogonal, unitary and symplectic
ensembles, respectively (abbreviated as GOE, GUE and GSE,
respectively) \cite{Meh91}.
For the one-dimensional $N$ Brownian motions, 
in which all particles start from the origin
and the nonintersecting condition is
imposed in the time interval $(0,T]$, we will show that
(i) at the very early stage, {\it i.e.} 
$t/T \ll 1$, the particle distribution is
described by using GUE, (ii) as time $t$ is going on, a
transition from GUE to GOE is observed, and (iii)
at the final stage $t=T$ the particle distribution can be
identified with GOE. 
As shown by the second equality of (\ref{eqn:Pbeta}),
the Gaussian ensemble of random matrices can be regarded
as the thermodynamical equilibrium of one dimensional 
gas system with (two-dimensional) Coulomb repulsive 
potential (\ref{eqn:W}) at the inverse temperature $\beta$. 
Here it should be noted that the vicious walkers
on a lattice have only contact repulsive interactions 
to satisfy the nonintersecting condition.
The global effective interactions among walkers
are accumulated by taking the diffusion scaling limit
and as its result a long-ranged Coulomb gas system is constructed.
Such {\it emergence of long-range effects in macroscopic scales
from systems having only short-ranged microscopic interactions}
is found only at critical points in thermodynamical equilibrium
systems, but it is a typical phenomenon observed in a various
interacting particle systems in far from equilibrium.

In particular, in the limit $T \to \infty$, that is, when
the nonintersecting condition will be imposed forever,
we can derive a system of stochastic differential equations
for the process with the drift terms which act as the
repulsive two-body forces proportional to the inverse of
distances between particles. In other words, the scaling
limit of vicious walks with $T \to \infty$ 
can realize Dyson's Brownian motion
model at $\beta=2$ \cite{Dys62}. 
It is reasonable to obtain such a stochastic process
from the vicious walker model, since it is known that Dyson's
Brownian motion model at $\beta=2$ can be mapped to the 
free fermion model \cite{Spohn,BF97}. The transition from
GUE to GOE is, however, first reported for vicious walkers 
with $T < \infty$ and explained using the two-matrix model
in the present paper.

\section{MODEL AND LINDSTR\"OM-GESSEL-VIENNOT DETERMINANT 
\label{sec:Model}}

One-dimensional vicious walks are defined as a subset 
of simple random walks as follows. 
Let $\{R_{k}^{s_{j}}\}_{k \geq 0}, j=1, 2, \cdots, N$, 
be the $N$ independent symmetric simple random walks on
${\bf Z}=\{ \cdots, -2, -1, 0, 1 , 2 \cdots \}$
started from $N$ distinct positions,
$2s_{1} < 2s_{2} < \cdots < 2s_{N}$,
$s_{j} \in {\bf Z}$. That is,
$$
R_{0}^{s_{j}}=2 s_{j}, \quad
\mbox{and} \quad
R_{k+1}^{s_{j}}=R_{k}^{s_{j}}-1 \quad
\mbox{or} \quad R_{k}^{s_{j}}+1,
$$
for $j=1,2, \cdots, N, k=0,1,2, \cdots.$
Fix the time interval $K$ as a positive even number.
The total number of walks is $2^{NK}$,
all of which are assumed to be realized with equal
probability $2^{-NK}$.
We consider a subset of walks such that any of walkers
does not meet other walkers up to time $K$.
In other words, the condition
\begin{equation}
R_{k}^{s_{1}} < R_{k}^{s_{2}} < \cdots
< R_{k}^{s_{N}}, \quad k=1,2, \cdots, K,
\label{eqn:vicious}
\end{equation}
is imposed.
Such a subset of walks is called the vicious walks
(up to time $K$) \cite{Fis84,HF84}.
Let $N_{N}(K; \{e_{j}\}| \{s_{j}\})$ be the total number
of the vicious walks, in which the $N$ walkers 
starting from $2s_{1} < 2 s_{2} < \cdots < 2 s_{N}$
arrive at the positions
$2e_{1} < 2e_{2} < \cdots < 2e_{N}$ at time $K$.
Then the probability that such vicious walks with
those fixed end-points are realized 
in all possible random walks started from
the given initial positions, 
which is denoted as $V_{N}(\{R_{k}^{s_{j}} \}_{k=0}^{K};
R_{K}^{s_{j}}=2 e_{j})$, is
$$
V_{N}(\{R_{k}^{s_{j}} \}_{k=0}^{K};
R_{K}^{s_{j}}=2 e_{j})=
\frac{N_{N}(K; \{e_{j}\} |\{s_{j}\})}{2^{NK}}.
$$
We also consider the probability
$$
V_{N}(\{R_{k}^{s_{j}} \}_{k=0}^{K}) 
=\sum_{e_{1} < e_{2} < \cdots < e_{N}}
V_{N}(\{R_{k}^{s_{j}} \}_{k=0}^{K};
R_{K}^{s_{j}}=2 e_{j}).
$$

Consider a subset of the square lattice ${\bf Z}^{2}$,
$$
{\cal L}_{K}=\{(x,y) \in {\bf Z}^{2}:
x+y=\mbox{even}, \ 0 \leq y \leq K \},
$$
and the set ${\cal E}_{K}$ of all edges which connect the
nearest-neighbor pairs of vertices in ${\cal L}_{K}$.
The lattice $({\cal L}_{K}, {\cal E}_{K})$ provides
the spatio-temporal plane and each walk of the $j$-th walker, 
$j=1,2, \cdots, N$, can be represented as a sequence of 
successive edges connecting vertices 
${\bf S}_{j}=(2s_{j}, 0)$ and ${\bf E}_{j}=(2e_{j}, K)$ 
on it, which we call the {\it lattice path}
running from ${\bf S}_{j}$ to ${\bf E}_{j}$.
If such lattice paths share a common vertex, they are said to 
intersect. Under the vicious walk condition
(\ref{eqn:vicious}), what we consider is a set of all 
$N$-tuples of {\it nonintersecting paths} \cite{Stem90}.
Let $\pi({\bf S} \to {\bf E})$ be
the set of all lattice paths from ${\bf S}$ to ${\bf E}$,
and
$\pi_{0}(\{{\bf S}_{j}\}_{j=1}^{N} 
\to \{{\bf E}_{j}\}_{j=1}^{N})$
be the set of all $N$-tuples $(\pi_{1}, \cdots, \pi_{N})$ of 
nonintersecting lattice
paths, in which $\pi_{j}$ runs from ${\bf S}_{j}$ to ${\bf E}_{j}$,
$j =1,2, \cdots, N$.
If we write the number of elements in a set $A$ as
$|A|$, then
$N_{N}(K;  \{e_{j}\} | \{s_{j}\})=
|\pi_{0}(\{{\bf S}_{j}\}_{j=1}^{N} \to \{{\bf E}_{j}\}_{j=1}^{N})|$.

The Lindstr\"om-Gessel-Viennot theorem
gives \cite{Lin73,GV85,Stem90} (see also \cite{Fis84,EG95,Fold}),
$$
N_{N}(K; \{e_{j}\} | \{s_{j}\})
= \det_{1 \leq j, k \leq N}
( |\pi({\bf S}_{k} \to {\bf E}_{j})|).
$$
Since
$
|\pi({\bf S}_{k} \to {\bf E}_{j})|=
{K \choose K/2+s_{k}-e_{j} },
$
we have the following binomial determinantal expressions
\begin{eqnarray}
&&V_{N}(\{R_{k}^{s_{j}}\}_{k=0}^{K};
R_{K}^{s_{j}}=2e_{j}) \nonumber\\
&& \quad= 2^{-NK} \det_{1 \leq j, k \leq N} 
\left( {K \choose K/2+s_{k}-e_{j}} \right),
\label{eqn:LGV1}
\end{eqnarray}
and
\begin{eqnarray}
&&V_{N}(\{R_{k}^{s_{j}}\}_{k=0}^{K}) \nonumber\\
&=& 2^{-NK} \sum_{e_{1} < e_{2} < \cdots < e_{N}}
\det_{1 \leq j, k \leq N} 
\left( {K \choose K/2+s_{k}-e_{j}} \right). \qquad
\label{eqn:LGV2}
\end{eqnarray}

\section{SCALING LIMIT OF VICISOUS WALKS \label{sec:Scaling}}

Recently Krattenthaler {\it et al.} \cite{KGV00} 
evaluated the asymptotes of (\ref{eqn:LGV2}) for large $K$ 
in the two special initial-configurations,
(i) $s_{j}=j-1$ and (ii) $s_{j}=2(j-1)$, as 
\begin{equation}
V_{N}(\{R_{k}^{s_{j}}\}_{k=0}^{K})
= a_{N} b_{N}(\{s_{j}\}) K^{-N(N-1)/4}
\left( 1+ {\cal O}(1/K) \right), 
\label{eqn:Krat1}
\end{equation}
where
\begin{eqnarray}
a_{N} &=&
\left\{
   \begin{array}{ll}
      (2^{N}/\pi)^{N/4} \
      \displaystyle{\prod_{j=1}^{N/2} (2j-2) ! }
      & \mbox{if} \ N=\mbox{even} \\
      (2^{N+1}/\pi)^{(N-1)/4}
      \displaystyle{\prod_{j=1}^{(N-1)/2} (2j-1) ! }
      & \mbox{if} \ N=\mbox{odd}, \\
   \end{array}\right.
\label{eqn:Krat2}
\end{eqnarray}
and
\begin{equation}
b_{N}(\{j-1\})=1, \quad b_{N}(\{2(j-1)\})=2^{N(N-1)/2}.
\label{eqn:Krat3}
\end{equation}

We found that their result can be immediately 
generalized as
$$
 b_{N}(\{s(j-1)\})=s^{N(N-1)/2} 
$$
for $s_{j}=s(j-1), s=1,2,3, \cdots.$
This observation suggests that we can take the 
scaling limit $L \to \infty$, where the time interval
$K \propto L$ and the initial spacing of walkers 
$s \propto \sqrt{L}$.

\subsection{Schur function}

In order to describe the scaling limit of the vicious walks,
the symmetric function called the Schur function is
useful. Here we give some of the fundamental
properties of Schur function \cite{Mac95,FH91,Ful97,Sta99}, 
which will be used below.

A partition 
$\lambda=(\lambda_{1}, \lambda_{2}, \cdots, \lambda_{N})$
is a non-increasing series of non-negative integers,
$\lambda_{1} \geq \lambda_{2} \geq \cdots \geq \lambda_{N}
\geq 0$.
Let $V$ be the $N$-dimensional complex vector space.
Then the Schur function 
$s_{\lambda}(z_{1}, \cdots, z_{N})$
associated with $\lambda$ is
a function of $(z_{1}, \cdots, z_{N}) \in V$
defined by 
\begin{equation}
s_{\lambda}(z_{1}, \cdots, z_{N}) =
\frac{\displaystyle{\det_{1 \leq j, k \leq N} 
\left(z_{j}^{\lambda_{k}+N-k}\right)}}
{\displaystyle{
\det_{1 \leq j, k \leq N} \left(z_{j}^{N-k}\right)}}.
\label{eqn:Schur}
\end{equation}
Let $\Delta_{\lambda}(\{z_{j}\})$ be the numerator
of (\ref{eqn:Schur}), which is an $N \times N$ determinant.
If we set $z_{\ell_{1}}=z_{\ell_{2}}$ for 
$1 \leq \ell_{1} < \ell_{2} \leq N$,
then $\Delta_{\lambda}(\{z_{j}\})=0$, since the
$\ell_{1}$-th row is equal to the $\ell_{2}$-th row.
Then it is divisible by each of
the differences $z_{\ell_{1}}-z_{\ell_{2}}, 
1 \leq \ell_{1} < \ell_{2} \leq N$,
and hence by their product
$\prod_{1 \leq j < k \leq N} (z_{j}-z_{k})$.
This product of all differences 
is known as the Vandermonde determinant,
which is nothing but the denominator of (\ref{eqn:Schur});
\begin{equation}
\Delta_{0}(\{z_{j}\}) \equiv
\det_{1 \leq j, k \leq N} (z_{j}^{N-k})
= \prod_{1 \leq j < k \leq N} (z_{j}-z_{k}).
\label{eqn:Vmd}
\end{equation}
Therefore it is concluded that
the ratio of two determinants
$\Delta_{\lambda}/\Delta_{0}$ is a polynomial
in $z_{1}, \cdots, z_{N}$. Moreover, it can be readily seen
from (\ref{eqn:Schur}) that the Schur function is 
a homogeneous polynomial of degree
$\sum_{j=1}^{N} \lambda_{j}$ in $z_{1}, \cdots, z_{N}$.

Let $q$ be a complex variable and set
$z_{j}=q^{j-1}$ in (\ref{eqn:Schur}). Then we have
$$
s_{\lambda}(1,q,q^2, \cdots, q^{N-1}) 
=
\frac{\displaystyle{\det_{1 \leq j, k \leq N} 
(q^{(j-1)(\lambda_{k}+N-k)})}}
{\displaystyle{\det_{1 \leq j, k \leq N} 
(q^{(j-1)(N-k)})}}. 
$$
Appropriate application of the formula of Vandermonde
determinant (\ref{eqn:Vmd}) gives the product form
\begin{eqnarray}
&&s_{\lambda}(1,q,q^2, \cdots, q^{N-1}) \nonumber\\
&& \quad
= q^{\sum(j-1) \lambda_{j}}
\prod_{1 \leq j < k \leq N}
\frac{q^{\lambda_{j}-\lambda_{k}+k-j}-1}
{q^{k-j}-1}.
\nonumber
\end{eqnarray}
Taking the limit $q \to 1$, we have the formula
\begin{equation}
s_{\lambda}(1,1, \cdots, 1)=
\prod_{1 \leq j < k \leq N}
\frac{\lambda_{j}-\lambda_{k}+k-j}{k-j}.
\label{eqn:Schur2}
\end{equation}
The Schur function is a character of the irreducible 
representation specified by $\lambda$ of the group
$GL(V)$ and (\ref{eqn:Schur2}) gives the dimension
of the representation.

\subsection{Diffusion scaling limit}

We set
\begin{equation}
K=Lt, \quad s_{j}=\frac{\sqrt{L}}{2} x_{j}, \quad
e_{j}=\frac{\sqrt{L}}{2} y_{j},  
\label{eqn:setting}
\end{equation}
for $j=1,2, \cdots, N,$
and take the limit $L \to \infty$.
Since in this limit each random walk $R_{k}^{s_{j}}$ 
converges to a Brownian motion, 
whose distribution function solves the diffusion equation,
this scaling limit is especially called 
{\it diffusion scaling limit}.
First we remark that,
for each strictly increasing series of integers
$y_{1} < y_{2} < \cdots < y_{N} $, 
a weakly decreasing series of integers
$\xi(y)=(\xi_{1}(y), \cdots, \xi_{N}(y))$ 
can be assigned by setting
\begin{equation}
 \xi_{j}(y)=y_{N-j+1}-(N-j), \
 j =1, 2, \cdots, N.
\label{eqn:xi}
\end{equation}
Then we can prove that,
for given $t >0$,
$x_{1} < x_{2} < \cdots < x_{N}$, and 
$y_{1} < y_{2} < \cdots < y_{N}$,
\begin{eqnarray}
&& \lim_{L \to \infty} \left(\frac{\sqrt{L}}{2}\right)^{N} 
V_{N} \left(
\left\{R_{k}^{\sqrt{L} x_{j} /2} \right\}_{k=0}^{Lt};
R_{Lt}^{\sqrt{L} x_{j}/2}=
\sqrt{L} y_{j} \right)
\nonumber\\
&=& (2 \pi t)^{-N/2} 
\det_{1 \leq j, k \leq N} \left[
\exp \left( - \frac{1}{2t} (x_{k}-y_{j})^2 \right) \right]
\nonumber\\
&=& (2 \pi t)^{-N/2} 
s_{\xi(y)}\left(e^{x_{1}/t}, e^{x_{2}/t}, \cdots,
e^{x_{N}/t}\right) \nonumber\\
&& \times \exp\left(- \frac{1}{2t} \sum_{j=1}^{N}
(x_{j}^2+y_{j}^2) \right)
h_{N}(\{e^{x_{j}/t}\}), 
\label{eqn:Lemma}
\end{eqnarray}
where $s_{\xi(y)}(z_{1}, \cdots, z_{N})$ is
the Schur function associated with $\xi(y)$,
defined by (\ref{eqn:Schur}) with $\lambda=\xi(y)$, and
\begin{eqnarray}
h_{N}(\{z_{j}\}) &\equiv&
\det_{1 \leq j, k \leq N} (z_{j}^{k-1}) \nonumber\\
&=& (-1)^{N(N-1)/2} \Delta_{0}(\{z_{j}\}) \nonumber\\
&=& \prod_{1 \leq j < k \leq N}(z_{k}-z_{j}).
\label{eqn:Vand}
\end{eqnarray}
The proof is given as follows.
Setting (\ref{eqn:setting}), we apply Stirling's formula 
to the RHS of (\ref{eqn:LGV1}) multiplied by
$(\sqrt{L}/2)^N$,
\begin{eqnarray}
&& \lim_{L \to \infty} 2^{-NLt} 
\left(\frac{\sqrt{L}}{2} \right)^{N}
\det_{1 \leq j, k \leq N}
\left( {Lt \choose 
\frac{Lt}{2}+\frac{\sqrt{L}(x_{k}-y_{j})}{2}} \right)
\nonumber\\
&=& 
\det_{1 \leq j, k \leq N}
\left( \lim_{L \to \infty} 2^{-Lt}
 \left(\frac{\sqrt{L}}{2} \right)  {Lt \choose 
\frac{Lt}{2}+\frac{\sqrt{L}(x_{k}-y_{j})}{2}} \right)
\nonumber\\
&=& 
\det_{1 \leq j, k \leq N}
\left[ \frac{1}{\sqrt{2 \pi t}} \ e^{-(x_{k}-y_{j})^2/2t} \right],
\label{eqn:Lemmab}
\end{eqnarray}
which gives the first equality of (\ref{eqn:Lemma}).
For the second equality, we rewrite (\ref{eqn:Lemmab}) as
$$
(2 \pi t)^{-N/2} e^{-\sum (x_{j}^2+y_{j}^2)/2t}
\det_{1 \leq j, k \leq N}
\left( e^{x_{k} y_{j}/t} \right). 
$$
The determinant is written as
\begin{eqnarray}
&& \det_{1 \leq j, k \leq N}
\left( e^{x_{k} y_{j}/t} \right) \nonumber\\
&& = \frac{
\det_{1 \leq j, k \leq N} \left(
(e^{x_{j}/t})^{y_{N-k+1}} \right)}
{\det_{1 \leq j, k \leq N} \left(
(e^{x_{j}/t})^{N-k} \right)}
\times h_{N} (\{ e^{x_{j}/t} \}). \nonumber
\end{eqnarray}
Using (\ref{eqn:xi}) and the definition of
Schur function (\ref{eqn:Schur}),
the second equality of (\ref{eqn:Lemma}) is obtained.

We consider the rescaled one-dimensional lattice
${\bf Z}/(\sqrt{L}/2)$, where the unit length is 
$2/\sqrt{L}$,
and let $\tilde{R}_{k}^{x}$ denote the symmetric simple
random walk starting from $x$ on ${\bf Z}/(\sqrt{L}/2)$.
Then (\ref{eqn:Lemma}) implies that
\begin{eqnarray}
&& \lim_{L \to \infty}
V_{N}\left(\left\{ \tilde{R}_{k}^{x_{j}} 
\right\}_{k=0}^{Lt};
\tilde{R}_{Lt}^{x_{j}} \in [y_{j}, y_{j}+dy_{j}]
\right) \nonumber\\
&& \qquad = f_{N}(t; \{y_{j}\}|\{x_{j}\}) d^{N} y.
\nonumber
\end{eqnarray}
Here we can give two expressions for
$f_{N}(t; \{y_{j}\}|\{x_{j}\})$,
\begin{eqnarray}
&& f_{N}(t; \{y_{j}\}|\{x_{j}\}) \nonumber\\
&=& (2 \pi t)^{-N/2} 
\det_{1 \leq j, k \leq N} \left[
\exp \left( - \frac{1}{2t} (x_{k}-y_{j})^2 \right) \right]
\nonumber\\
&=& (2 \pi t)^{-N/2} 
s_{\xi(y)}\left(e^{x_{1}/t}, e^{x_{2}/t}, \cdots,
e^{x_{N}/t}\right) \nonumber\\
&& \ \times \exp\left(- \frac{1}{2t} \sum_{j=1}^{N}
(x_{j}^2+y_{j}^2) \right)
h_{N}(\{e^{x_{j}/t}\}). 
\label{eqn:fN}
\end{eqnarray}

Since the vicious walkers are defined by imposing
the nonintersecting condition (\ref{eqn:vicious})
up to a given time $K$, the process depends on the
choice of $K$. That is, the process is 
{\it temporally inhomogeneous}.
This feature should be inherited in the process obtained 
in the diffusion scaling limit.
Since each random walk converges to a Brownian motion
in the diffusion scaling limit, the limit process of the
$N$ vicious walkers can be called the
{\it $N$ nonintersecting Brownian motions}.
Let $T > 0$ and we consider the $N$ nonintersecting
Brownian motions in the time interval $(0,T]$.
Set
$$
{\cal N}_{N}(t; \{x_{j}\}) =
\int_{y_{1} < \cdots < y_{N}} d^{N}y \
f_{N}(t; \{y_{j}\}|\{x_{j}\}).
$$
For $0 \leq s < t \leq T, \
x_{1} < \cdots < x_{N}, y_{1} < \cdots < y_{N}$,
the transition probability density from the
configuration $\{x_{j}\}$ at time $s$ to $\{y_{j}\}$
at $t$ is given by
\begin{equation}
g_{N}^{T}(s, \{x_{j}\}; t, \{y_{j}\}) =
\frac{f_{N}(t-s; \{y_{j}\}|\{x_{j}\}) 
{\cal N}_{N}(T-t; \{y_{j}\})}{{\cal N}_{N}(T-s; \{x_{j}\})},
\label{eqn:inhom}
\end{equation}
since the numerator in RHS gives the nonintersecting probability
for $(0,T]$ specified with the configurations $\{x_{j}\}$
and $\{y_{j}\}$ at times $s$ and $t$, respectively,
and the denominator gives the probability only specified
with $\{x_{j}\}$ at $s$,
where we have used the Markov property of the process.
The temporal inhomogeneity is obvious, since RHS depends not only
$t-s$ but also $T-s$ and $T-t$. 

\subsection{$t \to \infty$ asymptote of 
${\cal N}_{N}(t; \{x_{j}\})$}

It should be noted that, since ${\cal N}_{N}(t; \{x_{j}\})$
is the integral of $f_{N}(t; \{y_{j}\}|\{x_{j}\})$
over all possible end-positions $\{y_{j}\}$, it is the
probability that $N$ Brownian motions starting from
$\{x_{j}\}$ do not intersect up to time $t$.
Before studying the stochastic process defined by the 
transition probability density (\ref{eqn:inhom}),
here we assume $|{\bf x}| \equiv \sum_{j=1}^{N} |x_{j}| < \infty$
and evaluate the $t \to \infty$ asymptote
of ${\cal N}_{N}(t; \{x_{j}\})$.
In order to do that, the second expression of
$f_{N}(t; \{y_{j}\}|\{x_{j}\})$ in (\ref{eqn:fN}) 
will be useful,
\begin{eqnarray}
&&{\cal N}_{N}(t; \{x_{j}\}) 
= \frac{e^{-\sum x_{j}^2/2t}}{(2 \pi t)^{N/2}} 
h_{N}(\{e^{x_{j}/t}\})  \nonumber\\
&&  \times
\int_{y_{1} < \cdots < y_{N}} d^{N}y \
s_{\xi(y)}(e^{x_{1}/t}, \cdots, e^{x_{N}/t})
e^{-\sum y_{j}^2/2t}. \nonumber
\end{eqnarray}
By (\ref{eqn:Schur2}), (\ref{eqn:xi}) and (\ref{eqn:Vand}),
\begin{eqnarray}
&& \lim_{t \to \infty} s_{\xi(y)}(e^{x_{1}/t}, 
\cdots, e^{x_{N}/t})
= s_{\xi(y)}(1,1, \cdots, 1) \nonumber\\
&& = h_{N}(\{y_{j}\}) / \prod_{1 \leq j < k \leq N} (k-j), 
\nonumber
\end{eqnarray}
and
$$ 
\lim_{t \to \infty} t^{N(N-1)/2}
h_{N}(\{e^{x_{j}/t}\}) = h_{N}(\{x_{j}\}).
$$
We define
$$
b_{N}(\{x_{j}\})=h_{N}(\{x_{j}\})/ 
\prod_{1 \leq j < k \leq N}(k-j).
$$
Remark that this definition of $b_{N}(\{x\})$ is
consistent with (\ref{eqn:Krat3}).
Then we have
\begin{eqnarray}
&& {\cal N}_{N}(t; \{x_{j}\}) =
t^{-N^2/2} \frac{b_{N}(\{x_{j}\})}{(2 \pi)^{N/2}N !}
 \nonumber\\
&& \quad \times 
\int d^{N}y \ e^{-\sum y_{j}^{2}/2t} |h_{N}(\{y_{j}\})|
\times \left(1+ {\cal O}(1/t) \right)
\nonumber\\
&& = t^{-N(N-1)/4} \frac{b_{N}(\{x_{j}\})}{(2 \pi)^{N/2}N !}
 \nonumber\\
&& \quad \times 
\int d^{N}u \ e^{-\sum u_{j}^{2}/2} |h_{N}(\{u_{j}\})|
\times \left( 1+ {\cal O}(1/t) \right),
\nonumber
\end{eqnarray}
as $t$ tends to infinity,
where we have used the facts that with the absolute values
the product of differences $|h_{N}(\{y_{j}\})|$ 
is invariant under permutation 
of $y_{j}$, and $u_{j}=y_{j}/ \sqrt{t}$.
The last integral is the special case 
($\gamma=1/2$ and $a=1/2$) of 
\begin{eqnarray}
&&\int d^{N} u \ e^{- a \sum u_{j}^{2}}
\prod_{1 \leq j < k \leq N}
|u_{k}-u_{j}|^{2 \gamma} 
\nonumber\\
&=& (2 \pi)^{N/2}
(2a)^{-N(\gamma(N-1)+1)/2} 
\prod_{j=1}^{N} \frac{\Gamma(1+j \gamma)}{\Gamma(1+\gamma)},
\label{eqn:Mehta1}
\end{eqnarray}
which is found in Mehta \cite{Meh91} (eq.(17.6.7) on page 354), 
whose proof was given in \cite{Mac82} by use of
Selberg's integral \cite{Sel44}.
Here $\Gamma(x)$ is the Gamma function with the values
$\Gamma(3/2)=\sqrt{\pi}/2$ and
$
\prod_{j=1}^{N} \Gamma(1+j/2) = 2^{-N(N-1)/2} 
(\sqrt{\pi}/2)^{N} N ! \ a_{N},
$
where $a_{N}$ is given by (\ref{eqn:Krat2}).
Then we have 
\begin{eqnarray}
&&{\cal N}_{N}(t; \{x_{j}\}) \nonumber\\
&=&
t^{-\psi_{N}} 2^{-2 \psi_{N}} 
a_{N} b_{N}(\{x_{j}\}) \left( 1+ {\cal O}(1/t) \right)
\label{eqn:limN}
\end{eqnarray}
with 
\begin{equation}
\psi_{N}=\frac{1}{4}N(N-1),
\label{eqn:psi}
\end{equation}
as $t$ tends to infinity,
where $\psi_{N}$ is known as the critical exponent
of survival probability of vicious walkers 
\cite{Fis84,HF84,MB93a,MB93b}.
Since
$$
t^{-\psi_{N}}2^{-2\psi_{N}} b_{N}(\{x_{j}\})
=(Lt)^{-\psi_{N}}b_{N}(\{\sqrt{L}x_{j}/2\}),
$$
(\ref{eqn:limN}) suggests that the result (\ref{eqn:Krat1})
with (\ref{eqn:Krat2}) and (\ref{eqn:Krat3}) of
Krattenthaler {\it et al.} shall be generalized
for arbitrary initial positions of vicious walkers
on the lattice.

\section{GAUSSIAN RANDOM MATRIX ENSEMBLES AND
DYSON'S BROWNIAN MOTIONS \label{sec:Gaussian}}

In this section we study two special choices of $T$ ;
$T=t$ and $T \to \infty$.
We show that there is an interesting correspondence
between these choices of $T$ and the Gaussian ensembles
of random matrices. 
In order to see it we consider
the limit $|{\bf x}| \to 0$, where
$|{\bf x}| \equiv \sum_{j=1}^{N} |x_{j}|$.
It will be shown that the second expression of
$f_{N}(t; \{y_{j}\}|\{x_{j}\})$ in (\ref{eqn:fN}) is
useful for taking this limit.

\subsection{$T=t$ case and GOE}

Since the first expression in (\ref{eqn:fN}) gives 
$\lim_{t \to 0} f_{N}(t; \{y_{j}\}|\{x_{j}\})=\prod_{j=1}^{N} 
\delta(x_{j}-y_{j})$
with Dirac's delta functions, 
${\cal N}_{N}(0; \{x_{j}\})=1$ for any $\{x_{j}\}$.
Then setting $T=t$ makes (\ref{eqn:inhom}) depend
only on $t-s$. Set $s=0$ and use the second expression 
in (\ref{eqn:fN}) for 
$f_{N}(t; \{y_{j}\}|\{x_{j}\})$ and 
${\cal N}_{N}(t; \{x_{j}\})$.
By virtue of the Schur function (\ref{eqn:Schur2}),
for $t > 0$ and $|{\bf x}| \ll 1$, we have
\begin{eqnarray}
&&f_{N}(t; \{y_{j}\}|\{x_{j}\}) 
=(2 \pi t)^{-N/2} h_{N}(\{e^{x_{j}/t}\}) \nonumber\\
&& \quad \times 
s_{\xi(y)}(1, \cdots, 1) 
e^{-\sum y_{j}^2/2t} \left( 1+ {\cal O}(|x|) \right)
\nonumber\\
&& = \frac{t^{-N/2}}{(2 \pi)^{N/2}} 
e^{-\sum y_{j}^2/2t} h_{N}(\{y_{j}\}) \nonumber\\
&& \qquad \times
\prod_{1 \leq j < k \leq N} 
\frac{e^{x_{k}/t}-e^{x_{j}/t}}{k-j} \times
\left( 1+ {\cal O}(|x|) \right),
\nonumber
\end{eqnarray}
and
\begin{eqnarray}
&&{\cal N}_{N}(t; \{x_{j}\}) 
=(2 \pi t)^{-N/2} h_{N}(\{e^{x_{j}/t}\}) \nonumber\\
&& \times \int_{y_{1} < \cdots < y_{N}} d^{N} y \
s_{\xi(y)}(1, \cdots, 1) 
e^{-\sum y_{j}^2/2t} \left( 1+ {\cal O}(|x|) \right)
\nonumber\\
&& = \frac{t^{N(N-1)/4}}{(2 \pi)^{N/2} c_{N}}
\prod_{1 \leq j < k \leq N} 
\frac{e^{x_{k}/t}-e^{x_{j}/t}}{k-j} \
\left( 1+ {\cal O}(|x|) \right),
\nonumber
\end{eqnarray}
where the integral (\ref{eqn:Mehta1}) was used and
$$
c_{N} = \frac{2^{N(N-2)/2}}{\pi^{N/2} a_{N}} =
\left( 2^{N/2} \prod_{j=1}^{N} 
\Gamma(j/2) \right)^{-1}.
$$
Then (\ref{eqn:inhom}) gives
$$
g_{N}^{t}(0, \{0\}; t, \{y_{j}\})
= c_{N}t^{-\zeta_{N}} e^{-\sum y_{j}^2/2t}
h_{N}(\{y_{j}\})
$$
for $y_{1} < \cdots < y_{N}$ with 
$$
\zeta_{N}=\frac{1}{4} N(N+1).
$$ 
It means that
$$
g_{N}^{t}(0, \{0\}; t, \{y_{j}\})
= N ! \ g_{N}^{{\rm GOE}}(\{y_{i}\}; t)
$$
for $y_{1} < \cdots < y_{N}$,
where 
\begin{equation}
g_{N}^{{\rm GOE}}(\{y_{j}\}; \sigma^2)
=\frac{c_{N}}{N !} \sigma^{-2\zeta_{N}} 
\exp\left(-\frac{1}{2 \sigma^2} \sum_{j=1}^{N} y_{j}^2 \right)
h_{N}(\{y_{j}\})
\label{eqn:GOE}
\end{equation}
is the probability density function of eigenvalues 
$\{y_{j}\}$ of random matrices
in the Gaussian orthogonal ensemble
with variance $\sigma^2$ \cite{Meh91}.

\subsection{$T \to \infty$ limit and GUE}

Let
$$
p_{N}(s, \{x_{j}\}; t, \{y_{j}\}) \equiv
\lim_{T \to \infty} 
g_{N}^{T}(s, \{x_{j}\}; t, \{y_{j}\}).
$$
By use of (\ref{eqn:limN}) we can determine the
explicit form for any initial configuration 
$x_{1} < x_{2} < \cdots < x_{N}$ in this case as
\begin{equation}
p_{N}(0, \{x_{j}\}; t, \{y_{j}\}) =
\frac{h_{N}(\{y_{j}\})}{h_{N}(\{x_{j}\})}
f_{N}(t; \{y_{j}\}|\{x_{j}\}),
\label{eqn:pN}
\end{equation}
where $h_{N}$ is given by (\ref{eqn:Vand}).
Moreover, if we take the limit $|{\bf x}| \to 0$, we have
\begin{equation}
p_{N}(0, \{ 0 \}; t, \{y_{j}\})
= c^{\prime}_{N}
t^{-\zeta^{\prime}_{N}} e^{-\sum y_{j}^{2}/2t}
h_{N}(\{y_{j}\})^2,
\label{eqn:pN0}
\end{equation}
with 
$$
\zeta^{\prime}_{N}=\frac{N^2}{2}
\quad \mbox{and} \quad 
c^{\prime}_{N}=\left( ( 2\pi)^{N/2} \prod_{j=1}^{N} \Gamma(j)
\right)^{-1}.
$$
That is, we have the identity
$$
p_{N}(0, \{0\}; t, \{y_{j}\})
= N ! \ g_{N}^{{\rm GUE}}(\{y_{i}\}; t),
$$
for $y_{1} < \cdots < y_{N}$,
where 
\begin{equation}
g_{N}^{{\rm GUE}}(\{y_{j}\}; \sigma^2)
=\frac{c_{N}^{\prime}}{N !} \sigma^{-2 \zeta_{N}^{\prime}} 
\exp\left(-\frac{1}{2 \sigma^2} \sum_{j=1}^{N} y_{j}^2 \right)
h_{N}(\{y_{j}\})^2
\label{eqn:GUE}
\end{equation}
is the probability density function of eigenvalues 
$\{y_{j}\}$ of random matrices
in the Gaussian unitary ensemble
with variance $\sigma^2$ \cite{Meh91}.

In the case $T \to \infty$, the nonintersecting condition 
will be imposed forever, 
while in the case $T=t$, there will 
be no condition in the future. 
{\it The distributions
of particles at present depend on the condition
in the future.} 

By generalizing the calculation, which we did
in the case $T=t$, for arbitrary $T$ and comparing the
result with (\ref{eqn:pN0}), we have
\begin{eqnarray}
\frac{g_{N}^{T}(0, \{0\}; t, \{y_{j}\})}
{p_{N}(0, \{0 \}; t, \{y_{j}\})}
= \bar{c}_{N} T^{\psi_{N}} 
\frac{{\cal N}_{N}(T-t; \{y_{j}\})}
{h_{N}(\{y_{j}\})} \qquad
\label{eqn:Imhof}
\end{eqnarray}
for $y_{1} < \cdots < y_{N}$, with (\ref{eqn:psi}) and
$$
\bar{c}_{N}=\frac{c_{N}}{c^{\prime}_{N}}=
\pi^{N/2} \prod_{j=1}^{N} \frac{\Gamma(j)}
{\Gamma(j/2)}.
$$
When $N=2$, we can consider the process
of one variable $y=y_{2}-y_{1}$.
In this case $g_{2}^{T}$ and $p_{2}$ define the {\it Brownian meander}
and the {\it Bessel process}, respectively, both of which
are stochastic processes well-studied in probability theory \cite{RY98}.
The equality (\ref{eqn:Imhof}) can be regarded as the
multi-variable generalization of
Imhof's relation \cite{Imh84} between the Brownian meander
and the Bessel process.

\subsection{Dyson's Brownian motions}

In the limit $T \to \infty$ we have obtained the compact expression
(\ref{eqn:pN}) for any $x_{1} < \cdots < x_{N}$ and 
$y_{1} < \cdots < y_{N}$.
In this section, we show that a system of stochastic differential 
equations can be explicitly derived for (\ref{eqn:pN}).
Using it we will explain why we have the GUE
distribution. 

Let 
$$
E_{k}(\{x_{j}\})=\sum_{j=1; j \not= k}^{N}
\frac{1}{x_{k}-x_{j}} \quad
\mbox{for} \ k=1,2, \cdots N.
$$
It is easy to verify that
\begin{equation}
 E_{k}(\{x_{j}\})
=  \frac{\partial}{\partial x_{k}}
\log h_{N}(\{x_{j}\}), 
\label{eqn:formula1}
\end{equation}
for $k=1,2, \cdots, N$, and
\begin{equation}
\sum_{k=1}^{N}\left[ 
\frac{\partial}{\partial x_{k}} E_{k}(\{x_{j}\})
+\left( E_{k}(\{x_{j}\}) \right)^{2} \right] =0.
\label{eqn:formula2}
\end{equation}
Using these equalities, we can prove that 
$p_{N}(0, \{x_{j}\}; t, \{y_{j}\})$
solves the equation
\begin{equation}
\frac{\partial}{\partial t} u(t; \{x_{j}\}) =
\frac{1}{2} {\mit\Delta} u(t; \{x_{j}\})
+ \sum_{k=1}^{N} E_{k}(\{x_{j}\}) 
\frac{\partial}{\partial x_{k}} u(t; \{x_{j}\}),
\label{eqn:diff}
\end{equation}
where ${\mit\Delta}=\sum_{k=1}^{N} \partial^{2}/
\partial x_{k}^2$.
The proof is the following.
First we remark that 
the first expression in (\ref{eqn:fN}) states that
$f_{N}$ is a finite summation of the products of 
Gaussian kernels and thus it satisfies the diffusion equation
\cite{For89}.
Therefore
$$
\frac{\partial}{\partial t} p_{N}(0, \{x_{j}\}; t, \{y_{j}\})
= \frac{1}{2} \frac{h_{N}(\{y_{j}\})}{h_{N}(\{x_{j}\})}
 {\mit\Delta} f_{N}(t; \{y_{j}\}|\{x_{j}\}).
$$
Then we can find that, if $\{E_{k}(\{x_{j}\})\}$ satisfy
the equations
\begin{eqnarray}
&&\sum_{k=1}^{N} E_{k}(\{x_{j}\})
\frac{1}{h_{N}(\{x_{j}\})}
\frac{\partial}{\partial x_{k}}
f_{N}(t; \{y_{j}\}|\{x_{j}\}) \nonumber\\
&=& - \sum_{k=1}^{N} \left\{\frac{\partial}{\partial x_{k}}
\frac{1}{h_{N}(\{x_{j}\})} \right\}
\left\{\frac{\partial}{\partial x_{k}}
f_{N}(t; \{y_{j}\}|\{x_{j}\})\right\} \qquad
\label{eqn:eq1}
\end{eqnarray}
and
\begin{equation}
\sum_{k=1}^{N} E_{k}(\{x_{j}\}) 
\frac{\partial}{\partial x_{k}} 
\frac{1}{h_{N}(\{x_{j}\})}
= -\frac{1}{2} \sum_{k=1}^{N} 
\frac{\partial^2}{\partial x_{k}^2}
\frac{1}{h_{N}(\{x_{j}\})},
\label{eqn:eq2}
\end{equation}
(\ref{eqn:diff}) holds with
$u(t; \{x_{j}\})=p_{N}(0, \{x_{j}\}; t, \{y_{j}\})$.
It is easy to see that (\ref{eqn:eq1}) is satisfied if
(\ref{eqn:formula1})
holds for any $k=1,2, \cdots, N$.
Moreover, using (\ref{eqn:formula1}), we can reduce
(\ref{eqn:eq2}) to (\ref{eqn:formula2}).
Then the proof is completed.

The above result implies that the process defined 
in the limit $T \to \infty$ is the system of $N$ 
particles with positions $x_{1}(t), x_{2}(t), \cdots, x_{N}(t)$
at time $t$ on the real axis, whose time evolution is governed
by the stochastic differential equations
\begin{equation}
dx_{k}(t)=E_{k}(\{x_{j}(t)\})dt + dB_{k}(t), 
\label{eqn:sde1}
\end{equation}
$k=1,2, \cdots, N,$
where $\{B_{k}(t)\}_{k=1}^{N}$ are the independent standard
Brownian motions;
\begin{eqnarray}
&& B_{j}(0)=0, \quad \langle B_{j}(t) \rangle = 0, \nonumber\\
&&
\langle (B_{j}(t)-B_{j}(s))(B_{k}(t)-B_{k}(s)) \rangle
= |t-s| \delta_{jk} \nonumber
\end{eqnarray}
for any $t,s >0, j, k =1,2, \cdots, N$.
Because of the scaling property of Brownian motion,
$\sqrt{a} B_{j}(t)$ is equal to $B_{j}(a t)$
in distribution for arbitrary $a >0$. 
Then, if we set $t=2t^{\prime}$
and write $x_{k}(t)=\tilde{x}_{k}(t^{\prime})$,
(\ref{eqn:sde1}) are in the $\alpha=0,
\beta=2$ case of the equations
\begin{equation}
d\tilde{x}_{k}(t^{\prime})
=-\beta \frac{\partial}{\partial \tilde{x}_{k}}
 W^{\alpha}(\{\tilde{x}_{j}(t^{\prime})\}) 
dt^{\prime} + \sqrt{2} \ dB_{k}(t^{\prime}), 
\label{eqn:Dyson}
\end{equation}
$k=1,2, \cdots, N,$ with
$$
W^{\alpha}(\{\tilde{x}_{j}\})= \alpha \sum_{j=1}^{n}
\frac{\tilde{x}_{j}^{2}}{2}
- \sum_{1 \leq j < k \leq n} 
\log(\tilde{x}_{k}-\tilde{x}_{j}).
$$
When $\alpha=1$, (\ref{eqn:Dyson}) is known as the 
stochastic differential equations for the Dyson Brownian 
motions at the inverse temperature $\beta$ 
and the stationary distribution 
$\propto \exp(-\beta W^{1}(\{\tilde{x}_{j}\}))$ \cite{Dys62}.
If $\alpha=0$, the factor 
$\exp(-\beta \alpha \sum \tilde{x}_{j}^2/2)$
will be replaced by $\exp(-\sum \tilde{x}_{j}^2/4 t^{\prime})$
for finite $t^{\prime}$ and thus 
when $t^{\prime}=\sigma^2/ \beta$
we may have the Gaussian distribution
$ \propto
\exp(-(\beta/4 \sigma^2) \sum \tilde{x}_{j}^2) 
h_{N}(\{\tilde{x}_{j}\})^{\beta}$.
Setting $\beta=2$ gives the form (\ref{eqn:GUE}).

It should be noted that the system of diffusion equations 
describing the Dyson Brownian motions with $\beta=2$ 
can be mapped to the free fermion model \cite{Spohn,BF97}.

For general $T < \infty$, we will have the stochastic
differential equations
$$
dx_{k}(t) = E_{k}^{T}(\{x_{j}(t) \}) dt
+ d B_{k}(t),
$$
for $k=1,2, \cdots, N$, with
$$
E_{k}^{T}(\{x_{j}\})= \frac{\partial}{\partial x_{k}}
\ln {\cal N}_{N}(T-t; \{x_{j}\}).
$$

\section{TWO-MATRIX MODEL
\label{sec:Two-matrix}}

In Section \ref{sec:Scaling} we have constructed a system of 
nonintersecting Brownian motions in one dimension
as the diffusion scaling limit of vicious walks.
The obtained transition probability density
(\ref{eqn:inhom}) is temporally inhomogeneous
and the particle distribution depends not only
the observation time $t-s$ but also on the
time interval $T$, in which nonintersecting condition 
is imposed. In the case that all particles start from the origin
at time $s=0$, it was shown in Section \ref{sec:Gaussian}
that, (i) if $T=t$, it can be identified
with the eigenvalue distribution of random
matrices in GOE, and (ii) if $T \to \infty$, it becomes GUE.

For a fixed $T < \infty$, the above results are summarized
as follows. Consider the one dimensional $N$ Brownian motions
all starting from the origin at time $t=0$. We impose
the nonintersecting condition for the time interval
$(0, T]$. As the ratio $t/T \to 0$, the particle
distribution is asymptotically described by GUE.
On the other hand, at $t=T$, it can be identified with GOE.
This implies that as time $t$ is going on from 0 to $T$, there
occurs a transition of distribution from GUE to GOE.
In this section, we study this transition.

GUE is the ensemble of complex hermitian matrices
and GOE is that of real symmetric matrices.
The degrees of freedom are, when the matrix sizes are $N$,
$N^{2}$ and $N(N+1)/2$, respectively.
If we change the variables from these independent matrix 
elements to the eigenvalues and other mutually independent
variables, and then if we integrate the distribution functions
over all variables other than eigenvalues, we will have
the probability density functions for $N$ real eigenvalues
as (\ref{eqn:GUE}) and (\ref{eqn:GOE}) \cite{Meh91}.

Although the vicious walker model has no matrix structure
at all, here we show that its diffusion scaling limit,
nonintersecting Brownian motions, 
can be regarded as the reduction of a one-parameter family
of ensembles of matrix structures to a variable space
of eigenvalues.
The ``hidden structure" is not a single matrix but
a two-matrix model, in which a complex hermitian matrix
is coupled with a real symmetric matrix.

In the first subsection we will derive the two-matrix model
from the nonintersecting Brownian motions
and the transition from GUE to GOE will be discussed
in the second subsection. In the third subsection
we will show that the obtained two-matrix model
can be identified with the two-matrix model
of Pandey and Mehta \cite{PM83,MP83} 
by appropriate scale transformation of
matrix elements.

\subsection{From vicious walker model to
two-matrix model}

The generalized Imhof relation (\ref{eqn:Imhof}) with 
(\ref{eqn:pN0}) gives
\begin{eqnarray}
&&g_{N}^{T}(0, \{0\}; t, \{y_{j}\}) \nonumber\\
&\propto& e^{-\sum y_{j}^{2}/2t} h_{N}(\{y_{j}\}) \int d^{N} z \
{\rm sgn}(h_{N}(\{z_{j}\})) \nonumber\\
&& \times
\det_{1 \leq j, k \leq N} \left[
\exp\left(-\frac{1}{2(T-t)}(y_{j}-z_{k})^2 \right) \right], 
\label{eqn:gNT2}
\end{eqnarray}
where ${\rm sgn}(x)=x/|x|$.
The RHS is rewritten as
\begin{eqnarray}
&& h_{N}(\{y_{j}\}) \int d^{N}z \ 
{\rm sgn}(h_{N}(\{z_{j}\})) \nonumber\\
&& \quad \times
\det_{1 \leq j, k \leq N} \left[
\exp \left( - \frac{1}{2t} y_{j}^{2}
-\frac{1}{2(T-t)} (y_{j}-z_{k})^2 \right)\right] \nonumber\\
&=& h_{N}(\{y_{j}\}) \int d^{N}z \ {\rm sgn}(h_{N}(\{z_{j}\}))
e^{-\sum z_{j}^{2}/2T} \nonumber\\
&& \quad \times
\det_{1 \leq j, k \leq N} \left[
\exp\left( - \frac{T}{2t(T-t)} 
\left(y_{j}-\frac{t}{T}z_{k} \right)^2
\right) \right].
\nonumber
\end{eqnarray}
Setting $(t/T)z_{j}=a_{j}, j=1,2, \cdots, N$, we have
\begin{eqnarray}
&& g_{N}^{T}(0, \{0\}; t, \{y_{j}\}) \nonumber\\
&\propto& h_{N}(\{y_{j}\}) \int d^{N}a \ {\rm sgn}(h_{N}(\{a_{j}\})) 
\exp\left(-\frac{T}{2t^2} \sum_{j=1}^{N}a_{j}^{2} \right)
\nonumber\\
&& \quad \times
\det_{1 \leq j, k \leq N} \left[ 
\exp \left( -\frac{T}{2t(T-t)} (y_{j}-a_{k})^2 \right) \right].
\label{eqn:gNTa1}
\end{eqnarray}
Consider an ensemble of $N \times N$ real symmetric
matrices $\{A\}$ with an integration measure
$$
dA \equiv \prod_{1 \leq j \leq k \leq N} dA_{jk}.
$$
Let $\{a_{1}, a_{2}, \cdots, a_{N}\}$ be the eigenvalues
of the matrix $A$ and
$\{p_{1}, p_{2}, \cdots, p_{N(N-1)/2}\}$ be other mutually
independent variables. Then
$$
dA=J(\{a_{j}\}, \{p_{j}\}) \prod_{j=1}^{N} da_{j} \
\prod_{k=1}^{N(N-1)/2} dp_{k},
$$
where $J(\{a_{j}\}, \{p_{j}\})$ is the Jacobian
$$
J(\{a_{j}\}, \{p_{j}\})
=\left| \frac{\partial (A_{11}, A_{12}, \cdots, A_{NN})}
{\partial (a_{1}, \cdots, a_{N}, p_{1}, \cdots,
p_{N(N-1)/2}) } \right|.
$$
It is known that we can write
$$
J(\{a_{j}\},\{p_{k}\})=|h_{N}(\{a_{j}\})| f(\{p_{k}\}),
$$
where $f(\{p_{k}\})$ is independent of $a_{j}$'s \cite{Meh91}.
Therefore, for any function $G(\{a_{j}\})$ of $\{a_{1}, \cdots, a_{N}\}$,
we have the identity
\begin{equation}
\int dA \ G(\{a_{j}\})= c \int \prod_{j=1}^{N}da_{j} \
|h_{N}(\{a_{j}\})| G(\{a_{j}\})
\label{eqn:equality1}
\end{equation}
with
$$
c= \int \prod_{k=1}^{N(N-1)/2} dp_{k} \ f(\{p_{j}\}).
$$
Set
\begin{eqnarray}
G(\{a_{j}\}) &=& \frac{1}{|h_{N}(\{a_{j}\})|} 
{\rm sgn}(h_{N}(\{a_{j}\}))
\exp\left(-\frac{T}{2t^2} \sum_{j=1}^{N}a_{j}^2 \right)
\nonumber\\
&& \times
\det_{1 \leq j, k \leq N} \left[
\exp \left(-\frac{T}{2t(T-t)} (y_{j}-a_{k})^2 \right)\right].
\nonumber
\end{eqnarray}
Then using the formula (\ref{eqn:equality1}), (\ref{eqn:gNTa1})
becomes
\begin{eqnarray}
&& g_{N}^{T}(0, \{0\}; t, \{y_{j}\}) \nonumber\\
&\propto& h_{N}(\{y_{j}\}) \int dA \ \frac{1}{h_{N}(\{a_{j}\})}
\exp \left( -\frac{T}{2t^2} \sum_{j=1}^{N}a_{j}^2 \right)
\nonumber\\
&& \times 
\det_{1 \leq j, k \leq N} \left[
\exp \left( -\frac{T}{2t(T-t)} (y_{j}-a_{k})^2 \right) \right].
\label{eqn:gNTa2}
\end{eqnarray}

Next we use the following integral formula
\cite{HC57,IZ80,Meh81}; 
for $N \times N$ hermitian matrices $A$ and $B$
having eigenvalues $\{a_{1}, \cdots, a_{N}\}$ and
$\{b_{1}, \cdots, b_{N}\}$, respectively,
and for any constant $\gamma$,
\begin{eqnarray}
&& \int dU \ \exp\left[ \gamma \
{\rm tr} (A-U^{\dagger}BU)^2 \right] \nonumber\\
&\propto& \frac{1}{h_{N}(\{a_{j}\}) h_{N}(\{b_{j}\})}
\det_{1 \leq j, k \leq N} \left[
\exp \left( \gamma (a_{j}-b_{k})^2 \right) \right], \nonumber
\end{eqnarray}
where the integral is taken over the group
of unitary matrices $U$.
Then (\ref{eqn:gNTa2}) can be written as
\begin{eqnarray}
&& g_{N}^{T}(0, \{0\}; t, \{y_{j}\}) \nonumber\\
&\propto& h_{N}(\{y_{j}\})^2 \int dU \int dA \
\exp\left(-\frac{T}{2t^2} {\rm tr} A^2 \right)
\nonumber\\
&& \quad \times \exp \left(
-\frac{T}{2t(T-t)} {\rm tr} (U^{\dagger}YU-A)^2 \right),
\label{eqn:gNTa3}
\end{eqnarray}
where $Y$ is the $N \times N$ diagonal matrix such that
$
  Y_{jk}=y_{j} \delta_{jk}.
$
Since $U$ is a unitary matrix,
$H = U^{\dagger} Y U$ is an $N \times N$ complex hermitian matrix.
Then the integrand of (\ref{eqn:gNTa3}) can be regarded as
a weight for two matrices $H$ and $A$ given as
$$
\exp \left( -{\rm tr}
\left( \gamma_{H} H^2 - \gamma_{HA} H A 
+ \gamma_{A} A^2 \right)\right)
$$
with
\begin{equation}
\gamma_{H}=\frac{T}{2t(T-t)}, \
\gamma_{HA}=\frac{T}{t(T-t)}, \
\gamma_{A}=\frac{T^2}{2t^2(T-t)}.
\label{eqn:coeff}
\end{equation}

Consider an ensemble of $N \times N$ 
complex hermitian matrices $\{H\}$
with the integration measure
$$
dH = \prod_{1 \leq j \leq k \leq N} d {\rm Re}(H_{jk}) 
\prod_{1 \leq j < k \leq N} d {\rm Im}(H_{jk}).
$$
For each complex hermitian matrix $H$, 
let $\{y_{1}, \cdots, y_{N}\}$ be
a set of eigenvalues and $U$ be the $N \times N$ unitary
matrix such that
$H=U^{\dagger} Y U$ with $Y_{jk}=y_{j} \delta_{jk}$.
Then it is known that the integration measure $dH$ 
can be factorized into the product of the Haar measure
for unitary matrices $dU$ and an integration measure
for eigenvalues \cite{Meh91,Dei00}
$$
dH \propto dU \times h_{N}(\{y_{j}\})^{2} 
\prod_{j=1}^{N} dy_{j}.
$$
Now we introduce a two-matrix model, which consists of
an $N \times N$ real symmetric matrix $A$ and an 
$N \times N$ complex hermitian matrix $H$, 
with a probability density function 
$$
\mu_{N}(H,A) =
\frac{1}{Z_{N}} \exp \left(-{\rm tr}(
\gamma_{H}H^2-\gamma_{HA}HA + \gamma_{A} A^2 ) \right).
$$
Here $\gamma_{H}, \gamma_{HA}, \gamma_{A}$ 
are given as (\ref{eqn:coeff}) and
$Z_{N}$ is the partition function of the
two-matrix model,
$$
Z_{N}=\int dH \int dA \
\exp \left(-{\rm tr}(
\gamma_{H}H^2-\gamma_{HA}HA + \gamma_{A} A^2 ) \right).
$$
Then the relation 
$$
g_{N}^{T}(0, \{0\}; t, \{y_{j}\})
\propto h_{N}(\{y_{j}\})^2 \int dU \int dA \
\mu_{N}(U^{\dagger} Y U, A)
$$
is established.

\subsection{Transition from GUE to GOE}

Consider the Gaussian ensembles of real symmetric matrices
$\{A\}$ and complex hermitian matrices $\{H\}$
with sizes $N$ with the probability density functions 
$$
\nu_{N}(A) = C_{A} \exp \left(-\frac{1}{2 \sigma_{A}^2}
{\rm tr} A^2 \right), 
$$
and
$$
\tilde{\nu}_{N}(H) = C_{H} \exp
\left(-\frac{1}{2 \sigma_{H}^2}
{\rm tr} H^2 \right),
$$
respectively, where
$$
  \sigma_{A}^2= \frac{t^2}{T}, \quad
  \sigma_{H}^2= t \left( 1- \frac{t}{T} \right),
$$
and
$
C_{A}=2^{-N/2} (\pi \sigma_{A}^2)^{-\zeta_{N}}, 
C_{H}=2^{-N/2} (\pi \sigma_{H}^2)^{-\zeta_{N}^{\prime}}.
$
Then consider the convolution
$$
\hat{\mu}_{N}(H)= \int dA \ \nu_{N}(A) \tilde{\nu}_{N}(H-A).
$$
Since, for $1 \leq j, k \leq N$, 
$$
H_{jk}={\rm Re}(H_{jk})+ i \ {\rm Im}(H_{jk})
$$
with $i=\sqrt{-1}$, and
$$
   {\rm Re}(A_{jk})=A_{jk}, \quad
   {\rm Im}(A_{jk})=0,
$$
the convolution is also Gaussian distribution in the form
\begin{eqnarray}
&& \hat{\mu}_{N}(H) \nonumber\\
&\propto& 
\exp \left( - \sum_{j, k}
\left\{ \frac{\left( {\rm Re}(H_{jk}) \right)^2}
{2(\sigma_{H}^2+\sigma_{A}^2)}
+ \frac{\left( {\rm Im}(H_{jk}) \right)^2}{2\sigma_{H}^2} 
\right\} \right). \nonumber
\end{eqnarray}
Then (\ref{eqn:gNTa3}) gives
\begin{eqnarray}
&& g_{N}^{T}(0, \{0\}; t, \{y_{j}\}) 
\propto h_{N}(\{y_{j}\})^2 
\int dU \ \hat{\mu}_{N}(H) \nonumber\\
&\propto& h_{N}(\{y_{j}\})^2 \times \nonumber\\
&&  \int dU \
\exp \left( - \sum_{j, k} 
\left\{ \frac{\left( {\rm Re}(H_{jk}) \right)^2}
{2 \sigma_{{\rm Re}}^2} 
+\frac{\left( {\rm Im}(H_{jk}) \right)^2}{2 \sigma_{{\rm Im}}^2}
\right\} \right), \nonumber\\
\label{eqn:gNTa4}
\end{eqnarray}
where $H=U^{\dagger} Y U$ and
\begin{equation}
\sigma_{{\rm Re}}^2=t, \quad
\sigma_{{\rm Im}}^2=t \left(1-\frac{t}{T} \right).
\label{eqn:sigma}
\end{equation}

Now the transition from GUE to GOE is explicitly represented
by the time-dependent variances (\ref{eqn:sigma}).
With a fixed finite $T$, if $0 < t \ll T$,
$
\sigma_{\rm Re}^2= t \simeq \sigma_{\rm Im}^2.
$
Then the real and imaginary parts of complex hermitian
matrix elements are equally distributed as in GUE.
While $\sigma^2_{\rm Re}$ increases linearly in $t$,
$\sigma_{\rm Im}^{2}$ increases in time $t$ only up to
time $t=T/2$ and then decreases in time.
At time $t=T$, $\sigma_{\rm Im}^2=0$,
which implies that the imaginary parts of matrix elements
are zeros with probability one.
Then the distribution is identified with GOE.

\subsection{Pandey-Mehta's two-matrix model}

As an interpolation between GUE and GOE,
Pandey and Mehta introduced a family of Gaussian ensembles of 
hermitian matrices $\{H\}$ with one parameter $\alpha \in [0,1]$
\cite{PM83,MP83},
\begin{eqnarray}
&& \mu^{\rm PM}_{N}(H, \alpha) \nonumber\\
&=& C_{\rm PM} 
\exp \left(-\sum_{j, k} \left\{
\frac{\left({\rm Re}(H_{jk}) \right)^2}{4v^2} 
+ \frac{\left({\rm Im}(H_{jk}) \right)^2}{4 v^2 \alpha^2} 
\right\} \right), \qquad 
\label{eqn:PM}
\end{eqnarray}
where 
$v^2=\{2 (1+\alpha^2)\}^{-1}$
and
$
C_{\rm PM}=2^{-N/2} \alpha^{-N(N-1)/2} 
(2 \pi v^2)^{-N^2/2}.
$
Set
\begin{equation}
\kappa=\sqrt{ \frac{t(2T-t)}{T} }.
\label{eqn:kappa}
\end{equation}
Then, it is easy to see that, if
\begin{equation}
\alpha^2=1- \frac{t}{T},
\label{eqn:alpha}
\end{equation}
the equality 
\begin{equation}
\kappa^N \hat{\mu}_{N}(\kappa H)= \mu^{\rm PM}_{N}(H, \alpha)
\label{eqn:eqPM}
\end{equation}
is established.

For an even integer $N$ and an antisymmetric $N \times N$ matrix 
$B = (b_{jk})$ we put
\begin{eqnarray}
&& {\rm Pf}_{1\le j<k \le N}(b_{jk}) \nonumber\\
&=& \frac{1}{(N/2)!} \sum_{\sigma} {\rm sgn} (\sigma)
b_{\sigma(1)\sigma(2)}b_{\sigma(3)\sigma(4)}\dots 
b_{\sigma(N-1)\sigma(N)},
\nonumber
\end{eqnarray}
where the summation is extended over all permutations $\sigma$
of $(1,2,\dots,N)$ with restriction
$\sigma(2k-1)<\sigma(2k)$, $k=1,2,\dots,N/2$.
This expression is known as the {\it Pfaffian} \cite{Stem90}.
Pandey and Mehta showed that the probability density
function of eigenvalues $\{y_{j}\}$ of the complex
hermitian matrices, which are distributed following (\ref{eqn:PM}),
is given by
\begin{eqnarray}
g^{{\rm PM}}_{N}(\{y_{j}\}, \alpha) &=&
C_{N}(\alpha) \exp \left[ - \frac{1}{2} (1+\alpha^2) 
\sum_{j=1}^{N} y_{j}^{2} \right] \nonumber\\
&& \times h_{N}(\{y_{j}\})
{\rm Pf}_{1 \leq j < k \leq N} (F_{jk}),
\label{eqn:PMA}
\end{eqnarray}
where, setting
$$
  f(x) \equiv \left( 
  \frac{1-\alpha^4}{\pi \alpha^2} \right)^{1/2}
  \int_{0}^{x} \exp \left(
  - \frac{1-\alpha^4}{4 \alpha^2} y^2 \right) dy,
$$
if $N$ is even,
$
  F_{jk}=f(y_{j}-y_{k}), j, k = 1,2, \cdots, N,
$
and if $N$ is odd, we use above and in addition
$
  F_{j, N+1}=-F_{N+1, j}=1, j=1,2, \cdots, N,
  F_{N+1, N+1}=0,
$
and
$
C_{N}(\alpha)^{-1} =2^{3N/2} (1-\alpha^2)^{N(N-1)/4} 
(1+\alpha^2)^{-N(N+1)/4} \prod_{j=1}^{N} \Gamma(1+j/2)$
\cite{PM83,MP83}.
Then the relation (\ref{eqn:gNTa4}) and 
the equality (\ref{eqn:eqPM}) with
(\ref{eqn:kappa}) and (\ref{eqn:alpha}) 
imply the expression
\begin{eqnarray}
&& g_{N}^{T}(0, \{0\}; t, \{y_{j}\}) 
= c_{N} T^{N(N-1)/4} t^{-N^2/2} h_{N}(\{y_{j}\}) \nonumber\\
&& \times 
\exp \left(- \frac{1}{2t} \sum_{j=1}^{N} y_{j}^{2} \right)
{\rm Pf}_{1 \leq j < k \leq \hat{N}} 
(\tilde{F}_{jk}(T-t, \{y_{\ell}\})),
\nonumber\\
\label{eqn:eqPM2}
\end{eqnarray}
where
$$
\hat{N}= \left\{ \begin{array}{cl}
 N & \qquad \mbox{if $N$ is even} \\
N+1 & \qquad \mbox{if $N$ is odd}, \\
\end{array} \right.
$$
and
\begin{eqnarray}
&& \tilde{F}_{jk}(t, \{y_{\ell}\}) \nonumber\\
&=&
\left\{ \begin{array}{cl}
\displaystyle{
\frac{2}{\sqrt{\pi}}
{\rm Erf} \left( \frac{y_{k}-y_{j}}{2 \sqrt{t}} \right)}
& \mbox{if} \quad 1 \leq j , k \leq N, \\
  &  \\
1 & 
\mbox{if} \quad  1 \leq j \leq N, k=N+1, \\
-1 & 
\mbox{if} \quad j=N+1, 1 \leq k \leq N, \\
0 & 
\mbox{if} \quad j=k=N+1 \\
\end{array} \right.
\nonumber
\end{eqnarray}
with
$$
{\rm Erf}(x)= \int_{0}^{x} du \ e^{-u^2}.
$$
In order to derive (\ref{eqn:PMA}), Pandey and Mehta
performed integration over alternative variables
and then used the theory of Pfaffian \cite{PM83,MP83,Meh91}.
In Appendix, we will give the {\it integration version} of
Okada's minor-summation formula \cite{Oka89}.
It should be noted that using it (\ref{eqn:eqPM2})
can be readily obtained from (\ref{eqn:gNT2})
and this derivation provides another proof of
the equality (\ref{eqn:eqPM}).

\section{CONCLUDING REMARKS}

In the present paper we performed the diffusion scaling limit of
vicious walker model in one dimension and constructed the
nonintersecting Brownian motions for any finite number $N$ of
particles all starting from the origin. There the Schur function
plays an important role to represent the transition probability
density. We have shown that the spatial distribution
of particles depends not only the observation time $t$
but also on the time interval $T$ in which the nonintersecting
condition is imposed, and it can be described by use of the
probability density function of eigenvalues of $N \times N$ 
random matrices in Gaussian ensembles. 
It was shown that the particle distribution
depends on the ratio $t/T$ and a transition from GUE 
distribution to GOE distribution occurs in its time development.
Such a transition between different ensembles of random matrices
had been studied in the two-matrix model by Pandey and Mehta,
in which a parameter $\alpha$ was introduced and a one-parameter family
of random matrix ensembles was considered.
The present work showed that the scaling limit of vicious walk
model realizes such a two-matrix model as a stochastic process,
in such a way that the parameter $\alpha$ is continuously changed 
following (\ref{eqn:alpha}) as the system is developing in time $t$
up to $T$.

In the present paper, since we have considered only
the transition probability density between two different times,
a two-matrix model was analyzed. 
As briefly reported in \cite{NKT02}, however, multi-time correlations
among particles at intermediate times 
between $t=0$ and $t=T$ can be identified with
inter-matrix correlations in a multimatrix model.
The corresponding multimatrix model for calculating 
$M$-intermediate-time correlation functions
is a version of Nagao's $(M+1)$ matrix model, 
where one real symmetric matrix is combined at the end of 
a chain of $M$ complex hermitian matrices \cite{Nag01}. 
This observation implies that the diffusion scaling limit
of the vicious walker model is mathematically identified with
a matrix chain, which is set along the time axis.
The time development from GUE to GOE in the process
can be then regarded as appearance of an {\it edge effect}, as the
observation time $t$ on the time axis is approaching to
the end point $t=T$ in this chain structure.
Further study on the relations between multimatrix models
and nonequilibrium interacting particle systems is desired.

\begin{acknowledgments}
The authors acknowledge useful discussions with
P. Forrester, T. Nagao, C. Krattenthaler, 
J. Cardy and T. Fukui.
\end{acknowledgments}

\appendix*
\section{INTEGRATION VERSION 
OF OKADA'S MINOR-SUMMATION FORMULA}

Assume $1 \leq n \leq m$ and let
$z(j,k), 1 \leq j \leq n, 1 \leq k \leq m$, be
indeterminates and $Z(z(j,k))$ be the
$n \times m$ matrix with $(j,k)$-element $z(j,k)$.
We consider the sum of all minors of $Z$ with
a given size $r$. That is,
we define
$$
d(a_{1}, \cdots, a_{r})
= \sum_{1 \leq b_{1} < b_{2} < \cdots < b_{r} \leq m}
\det_{1 \leq j, k \leq r} (z(a_{j}, b_{k}))
$$
for $r = 1,2, \cdots, m$. 
Okada proved the following equalities known as the
{\it minor-summation formula} \cite{Oka89};
if $r$ is odd, then
\begin{eqnarray}
&& d(a_{1}, \cdots, a_{r}) = \nonumber\\
&&{\rm Pf}
\left( \matrix{
0 & d(a_{1}) & d(a_{2}) & \cdots & d(a_{r}) \cr
-d(a_{1}) & 0 & d(a_{1}, a_{2}) & \cdots
& d(a_{1}, a_{r}) \cr
-d(a_{2}) & -d(a_{1}, a_{2}) & 0 &
\cdots & d(a_{2}, a_{r}) \cr
      &   &  \cdots & & \cr
-d(a_{r}) & -d(a_{1}, a_{r}) & -d(a_{2}, a_{r}) & 
\cdots & 0 
} \right),
\nonumber
\end{eqnarray}
if $r$ is even, then
\begin{eqnarray}
&& d(a_{1}, \cdots, a_{r}) = \nonumber\\
&& {\rm Pf}
\left( \matrix{
 0 & d(a_{1}, a_{2}) & d(a_{1}, a_{3}) & \cdots
& d(a_{1}, a_{r}) \cr
 -d(a_{1}, a_{2}) & 0 & d(a_{2}, a_{3}) &
\cdots & d(a_{2}, a_{r}) \cr
 -d(a_{1}, a_{3}) & -d(a_{2}, a_{3}) & 0 &
\cdots & d(a_{3}, a_{r}) \cr
   &  \cdots & & \cr
-d(a_{1}, a_{r}) & -d(a_{2}, a_{r}) & 
-d(a_{3}, a_{r}) &
\cdots & 0 
} \right). \nonumber
\end{eqnarray}

Now we give the integration version of Okada's formula.
Let $z(x,y)$ be a square integrable continuous function of
real variables $x, y$.
Then
\begin{eqnarray}
&& \int_{-\infty < y_{1} < \cdots < y_{n} < \infty}
d^{n}y \det_{1 \le j,k \le n}(z(x_j, y_k)) \nonumber\\
&=& {\rm Pf}_{1 \leq j < k \leq \hat{n}}( F_{jk}(\{x_{\ell}\}) ),
\label{eqn:cont}
\end{eqnarray}
where
$$
\hat{n}= \left\{ \begin{array}{cl}
 n & \qquad \mbox{if $n$ is even} \\
n+1 & \qquad \mbox{if $n$ is odd}, \\
\end{array} \right.
$$
and 
$$
F_{jk}(\{x_{\ell}\}) =
\left\{ \begin{array}{cl}
I_z(x_{j}, x_{k}) 
& \qquad \mbox{if} \quad 1 \leq j < k \leq n, \\
-I_z(x_{j}, x_{k}) 
& \qquad \mbox{if} \quad 1 \leq k < j \leq n, \\
I_z(x_j) & \qquad
\mbox{if} \quad 1 \leq j \leq n, k=n+1, \\
-I_z(x_k) & \qquad
\mbox{if} \quad j=n+1, 1 \leq k \leq n, \\
0 & \qquad
\mbox{if} \quad 1 \leq j=k \leq n+1 \\
\end{array} \right.
$$
with
\begin{eqnarray}
&&I_z(x_{j}) = \int_{-\infty}^{\infty} z(x_{j},y )dy,
\nonumber
\\
&&I_z (x_{j}, x_{k}) = 
\int_{- \infty < y_{1} < y_{2} < \infty} 
\left| \matrix{
z(x_{j},y_{1}) & z(x_{j},y_{2}) \cr
z(x_{k},y_{1}) & z(x_{k},y_{2}) } \right|
dy_{1} dy_{2}.
\nonumber
\end{eqnarray}

The proof is the following.
We write the integral in LHS of 
(\ref{eqn:cont}) as a limit of
summation;
\begin{eqnarray}
&&\int_{- \infty < y_{1} < \cdots < y_{n} < \infty}
d^n y \det_{1 \leq j, k \leq n}
\left( z(x_{j},y_{k})\right) \nonumber\\
&=&\lim_{M \to \infty}
\int_{- M/2 \leq y_{1} < \cdots < y_{n} \leq M/2}
d^n y \det_{1 \leq j, k \leq n}
\left( z(x_{j},y_{k})\right)
\nonumber\\
&=& \lim_{M \to \infty} \lim_{\delta \to 0} \
\delta^{n} 
\sum_{1 \leq b_{1} < \cdots < b_{N}
\leq m(M, \delta)} 
\det_{1 \leq j,k \leq n} (z(x_{j}, \hat{y}(b_{k}) )),
\nonumber
\end{eqnarray}
where
$ m(M,\delta) = [M/\delta]$, the greatest integer
not greater than $M/\delta$,
and
$$
\hat{y}(b)=\frac{M-\delta}{m(M,\delta)-1} b
- \left\{ \frac{M-\delta}{m(M, \delta)-1}
+\frac{M}{2} \right\}.
$$

Let
\begin{eqnarray}
\tilde{d}(x_{j}) &=&
\sum_{1 \leq b_{1}\leq m(M, \delta)}
z(x_{j}, \hat{y}(b_{1})) ,
\nonumber
\\
\tilde{d}(x_{j}, x_{k}) &=&
\sum_{1 \leq b_{1} < b_{2} \leq m(M, \delta)}
\left| \matrix{
z(x_{j}, \hat{y}(b_{1})) & z(x_{j}, \hat{y}(b_{2})) \cr
z(x_{k}, \hat{y}(b_{1})) & z(x_{k}, \hat{y}(b_{2})) } \right|,
\nonumber
\end{eqnarray}
and set
$$
S_{jk}(\{x_{\ell}\})= 
\left\{ \begin{array}{cl}
\tilde{d}(x_{j}, x_{k}) &
\mbox{if} \ 1 \leq j < k \leq n, 
\\
-\tilde{d}(x_{j}, x_{k}) &
\mbox{if} \ 1 \leq k < j \leq n, 
\\
\tilde{d}(x_{j}) & 
\mbox{if} \ 1 \leq j \leq n, k=n+1, 
\\
- \tilde{d}(x_{j}) & 
\mbox{if} \ j=n+1, 1 \leq k \leq n, 
\\
0 & \mbox{if} \ 1 \leq j=k \leq n+1. 
\\
\end{array}\right.
$$
Then Okada's formula gives
\begin{eqnarray}
&&\int_{-\infty < y_{1} < \cdots < y_{n} < \infty}
d^n y \det_{1 \le j,k \le n}(z(x_j, y_k)) \nonumber\\
&=& \lim_{M \to \infty} \lim_{\delta \to 0} \ \delta^{n} \ 
{\rm Pf}_{1 \leq j < k \leq \hat{n}}( S_{jk}(\{x_{\ell}\}) ).
\label{eqn:PfA}
\end{eqnarray}

Since the Pfaffian in (\ref{eqn:PfA}) is
a finite summation of
$n/2$ products of $\tilde{d}(x_{j}, x_{k})$'s if $n$ is even, and
it is a finite summation
of $(n-1)/2$ products of $\tilde{d}(x_{j}, x_{k})$'s 
multiplied by $\tilde{d}(x_{\ell})$ if $n$ is odd, we may have 
\begin{eqnarray}
&& \int_{-\infty < y_{1} < \cdots < y_{n} < \infty}
 d^{n} y \det_{1 \le j,k \le n}(z(x_j, y_k))
\nonumber\\\
&=&
{\rm Pf}_{1 \leq j < k \leq \hat{n}} \left(
\lim_{M \to \infty} \lim_{\delta \to 0} \
\delta^{\alpha(j,k)} \ S_{jk}(\{x_{\ell}\}) \right),
\nonumber
\end{eqnarray}
where
\begin{eqnarray}
\alpha(j,k)&=& 2 \qquad \mbox{for} \ 1 \leq j < k \leq n,
\nonumber\\
\alpha(j, n+1)&=& 1 \qquad \mbox{for} \ 1 \leq j \leq n.
\nonumber
\end{eqnarray}
Since $z(x,y)$ is assumed to be square integrable and continuous, 
\begin{eqnarray}
\lim_{M \to \infty} \lim_{\delta \to 0}
\delta S_{j n+1}(\{x_{\ell}\}) &=& I_z(x_j),
\nonumber
\\
\lim_{M \to \infty} \lim_{\delta \to 0}
\delta^2 S_{jk}(\{x_{\ell}\})&=& I_z(x_j,x_k), \nonumber
\end{eqnarray}
for $ 1 \leq j,k \leq n$.
Then the proof is completed.

By elementary calculation we can show that
\begin{eqnarray}
&&I_{z} (x)=1, \nonumber\\
&&I_{z} (x,y)=
\frac{2}{\sqrt{\pi}} {\rm Erf} \left( \frac{y-x}{2\sqrt{t}}\right)
\nonumber
\end{eqnarray}
for
$
  z(x,y)=\frac{1}{\sqrt{2 \pi t}} e^{-(x-y)^2/2t}.
$
Applying (\ref{eqn:cont}) with $n=N$,
the expression (\ref{eqn:eqPM2}) is obtained
from (\ref{eqn:gNT2}).


\bibliography{basename of .bib file}

\begin{thebibliography}{99}
\bibitem{Fis84}
M. E. Fisher, J. Stat. Phys. {\bf 34}, 667 (1984).

\bibitem{GOV98}
A. J. Guttmann, A. L. Owczarek and X. G. Viennot,
J. Phys. A {\bf 31}, 8123 (1998).

\bibitem{KGV00}
C. Krattenthaler, A. J. Guttmann, and
X. G. Viennot, 
J. Phys. A {\bf 33}, 8835 (2000).

\bibitem{HF84}
D. A. Huse and M. E. Fisher,
Phys. Rev. B {\bf 29}, 239 (1984).

\bibitem{AME91}
D. K. Arrowsmith, P. Mason, and J. W. Essam, 
Physica A {\bf 177}, 267 (1991).

\bibitem{EG95}
J. W. Essam and A. J. Guttmann,
Phys. Rev. E {\bf 52}, 5849 (1995).

\bibitem{Mac95}
I. G. Macdonald,
{\it Symmetric Functions and Hall Polynomials},
2nd ed, (Oxford Univ. Press, Oxford, 1995).

\bibitem{Ful97}
W. Fulton,
{\it Young Tableaux},
(Cambridge Univ. Press, Cambridge, 1997).

\bibitem{Sta99}
R. P. Stanley,
{\it Enumerative Combinatorics}, vol.2,
(Cambridge Univ. Press, Cambridge, 1999).

\bibitem{FH91}
W. Fulton and J. Harris, 
{\it Representation Theory},
(Springer, New York, 1991).

\bibitem{Joh00}
K. Johansson,
Commun. Math. Phys. {\bf 209}, 437 (2000).

\bibitem{Bai00}
J. Baik,
Commun. Pure Appl. Math.
{\bf 53}, 1385 (2000).

\bibitem{NF02}
T. Nagao and P. J. Forrester,
Nucl. Phys. {\bf B620}[FS], 551 (2002).

\bibitem{Fold}
See also the earlier works on vicious walks by
P. J. Forrester, J. Phys. A {\bf 22}, L609 (1989);
A {\bf 23}, 1259 (1990); A {\bf 24}, 203 (1991).

\bibitem{Meh91}
M. L. Mehta,
{\it Random Matrices}, 2nd ed,
(Academic, New York, 1991).

\bibitem{Dei00}
P. Deift,
{\it Orthogonal Polynomials and Random Matrices:
A Riemann-Hilbert Approach},
(American Math. Society,
Providence, 2000).

\bibitem{Spohn}
On a torus another construction of nonintersecting Brownian
motions and Dyson's model was given by
H. Spohn, in {\it Hydrodynamic Behavior of Interacting
Particle Systems}, ed. G. Papanicolaou,
(Springer, Berlin, 1987);
H. Spohn, 
Markov Processes and Related Fields {\bf 4}, 649 (1998).

\bibitem{Lin73}
B. Lindstr\"om, Bull. London Math. Soc.
{\bf 5}, 85 (1973).

\bibitem{GV85}
I. Gessel and X. G. Viennot,
Adv. Math. {\bf 58}, 300 (1985).

\bibitem{Stem90}
J. R. Stembridge,
Adv. Math. {\bf 83}, 96 (1990).

\bibitem{PM83}
A. Pandey and M. L. Mehta,
Commun. Math. Phys. {\bf 87}, 449 (1983).

\bibitem{MP83}
M. L. Mehta and A. Pandey,
J. Phys. A {\bf 16}, 2655 (1983).

\bibitem{Dys62}
F. J. Dyson, J. Math. Phys. {\bf 3}, 1191 (1962).

\bibitem{BF97}
T. H. Baker and P. J. Forrester,
Commun. Math. Phys. {\bf 188}, 175 (1997).

\bibitem{Mac82}
I. G. Macdonald, 
SIAM J. Math. Anal. {\bf 13}, 988 (1982).

\bibitem{Sel44}
A. Selberg, 
Norsk Mat. Tidsskrift {\bf 26}, 71 (1944).

\bibitem{MB93a}
S. Mukherji and S. M. Bhattacharjee,
J. Phys. A {\bf 26}, L1139 (1993).

\bibitem{MB93b}
S. Mukherji and S. M. Bhattacharjee,
Phys. Rev. E {\bf 48}, 3427 (1993); 
E {\bf 52}, 3301(E) (1995).

\bibitem{RY98}
D. Revuz and M. Yor,
{\it Continuous Martingales and Brownian Motion},
3rd edition, (Springer, New York, 1998).

\bibitem{Imh84}
J. -P. Imhof,
J. Appl. Prob. {\bf 21}, 500 (1984).

\bibitem{For89}
P. J. Forrester,
J. Stat. Phys. {\bf 56}, 767 (1989);
{\bf 76}, 331 (1994).

\bibitem{HC57}
Harish-Chandra, Am. J. Math. {\bf 79}, 87 (1957)

\bibitem{IZ80}
C. Itzykson and J. -B. Zuber, J. Math. Phys. 
{\bf 21}, 411 (1980).

\bibitem{Meh81}
M. L. Mehta, Commun. Math. Phys. {\bf 79}, 327 (1981).

\bibitem{Oka89}
S. Okada,
J. Combin. Theor, Ser. A {\bf 51}, 1 (1989).

\bibitem{NKT02}
T. Nagao, M. Katori and H. Tanemura, 
cond-mat/0202068.

\bibitem{Nag01}
T. Nagao, Nucl. Phys. {\bf B602}, 622 (2001).

\end{thebibliography}

\end{document}